\documentclass[prl,aps,reprint,superscriptaddress,showpacs]{revtex4-1}
\usepackage{latexsym,amsmath,amssymb,setspace}
\usepackage{graphicx}
\usepackage{dcolumn}
\usepackage{bm}

\newif\ifAMStwofonts

\ifAMStwofonts \else 

\fi

\newcommand{\beq}{\begin{equation}}
\newcommand{\eeq}{\end{equation}}
\newcommand{\bdm}{\begin{displaymath}}
\newcommand{\edm}{\end{displaymath}}

\newcommand{\ea}{et al.}

\newcommand{\eqn}[1]{eq.~({#1})}

\newcommand{\cc}{$\rm{cm}^{3}$}

\begin{document}

\title{Spontaneous Transition of Turbulent Flames to Detonations in
Unconfined Media}



\author{Alexei Y. Poludnenko}
\email[Corresponding author: ]{apol@lcp.nrl.navy.mil}
\affiliation{Naval Research Laboratory, Washington, D.C. 20375, USA}

\author{Thomas A. Gardiner}
\affiliation{Sandia National Laboratories, Albuquerque, N.M. 87185-1189,
USA}

\author{Elaine S. Oran}
\affiliation{Naval Research Laboratory, Washington, D.C. 20375, USA}

\date{\today}

\begin{abstract}
Deflagration-to-detonation transition (DDT) can occur in environments
ranging from experimental and industrial systems to astrophysical
thermonuclear (type Ia) supernovae explosions.  Substantial progress
has been made in explaining the nature of DDT in confined systems with
walls, internal obstacles, or pre-existing shocks. It remains unclear,
however, whether DDT can occur in unconfined media. Here we use direct
numerical simulations (DNS) to show that for high enough turbulent
intensities unconfined, subsonic, premixed, turbulent flames are
inherently unstable to DDT. The associated mechanism, based on the
nonsteady evolution of flames faster than the Chapman-Jouguet
deflagrations, is qualitatively different from the traditionally
suggested spontaneous reaction wave model, and thus does not require
the formation of distributed flames. Critical turbulent flame speeds,
predicted by this mechanism for the onset of DDT, are in agreement
with DNS results.
\end{abstract}


\maketitle

Since the discovery of detonations, the question of the physical
mechanisms that create these self-supporting, supersonic, shock-driven
reaction waves has been a forefront topic in combustion
theory. Uncontrolled development of detonations poses significant
threats to chemical storage and processing facilities, mining
operations, etc. \citep{Nettleton}, while controlled detonation
initiation in propulsion systems could revolutionize transportation
\citep{Roy04}. On astrophysical scales, detonation formation is
presently the most important, yet least understood, aspect of the
explosion \citep{Khokhlov91,Gamezo04} powering type Ia supernovae,
which, as standard cosmological distance indicators, led to the
discovery of the accelerating expansion of the Universe
\citep{Riess98,Perlmutter99}.

Early studies \citep{Mallard} showed that a detonation can arise from
a slow, highly subsonic deflagration ignited in an initially
unpressurized system. Significant progress has since been made
experimentally \citep{Urtiew,*Kuznetsov} and numerically
\citep{Kagan2003,OranGamezo,Gamezo08,*Kessler10,Bychkov08,*Liberman10} in
elucidating the physics of the deflagration-to-detonation transition
(DDT) in confined systems, and particularly in closed channels. These
studies showed that the confining effect of channel walls on the hot,
expanding products of burning and the interaction of the resulting
flow with walls and obstacles are important in accelerating the flame
and causing the pressure increase, thus creating conditions necessary
for the detonation ignition. This raises the question: Is DDT possible
in unconfined media without assistance of walls or obstacles, e.g., in
unconfined clouds of fuel vapor or in the interior of a white dwarf
star during a supernova explosion?

Zel'dovich \ea~\citep{Zeldovich} originally suggested that a
detonation can form in a region (``hot spot'') with a suitable
gradient of reactivity. The resulting spontaneous reaction wave
propagating through that gradient creates a pressure wave that can
eventually develop into a shock and then a detonation
\citep{Kapila,Khokhlov97ApJ}. In confined systems, multidimensional
direct numerical simulations (DNS) have shown that hot spots can form
through repeated shock-flame interactions and fuel compression by
shocks \citep{OranGamezo}.

It remains unclear, however, if and how hot spots could form in
unconfined, unpressurized media. The most likely mechanism involves
flame interactions with intense turbulence. It was suggested
\citep{Khokhlov97ApJ,Niemeyer97} that the flame structure could be
disrupted by turbulence, producing a distributed flame with reactivity
gradients capable of initiating a detonation. There are, however, no
realistic \textit{ab initio} experimental or numerical demonstrations
of this process. Here we show that high-speed turbulence-flame
interactions can indeed lead to DDT, but through a different process
in which pressure build-up in the system does not rely on the
propagation of global spontaneous reaction waves and, thus, does not
require the formation of large-scale gradients of reactivity.

\textit{Model and method.} ---  The DNS presented here solve the
compressible reactive-flow equations including thermal conduction,
molecular species diffusion, and energy release
\citep{Poludnenko10,Poludnenko11}. They use an ideal-gas
equation of state and a single-step, first-order Arrhenius kinetics to
describe chemical reactions converting fuel into product. Simplified
reaction-diffusion models represent stoichiometric H$_2$-air and
CH$_4$-air mixtures with unity Lewis number and reproduce both
experimental laminar flame and detonation properties
\citep{Gamezo08,*Kessler10}. Simulations were performed with the code
Athena-RFX, which uses a fully unsplit corner-transport upwind scheme
with PPM spatial reconstruction and the HLLC Riemann solver
\citep{Gardiner08,Athena,Poludnenko10,Poludnenko11}. Turbulence is driven
using a spectral method \citep{Lemaster09,Poludnenko10}.



\begin{figure}[!t]
\quad \quad \includegraphics[clip, width=0.435\textwidth]{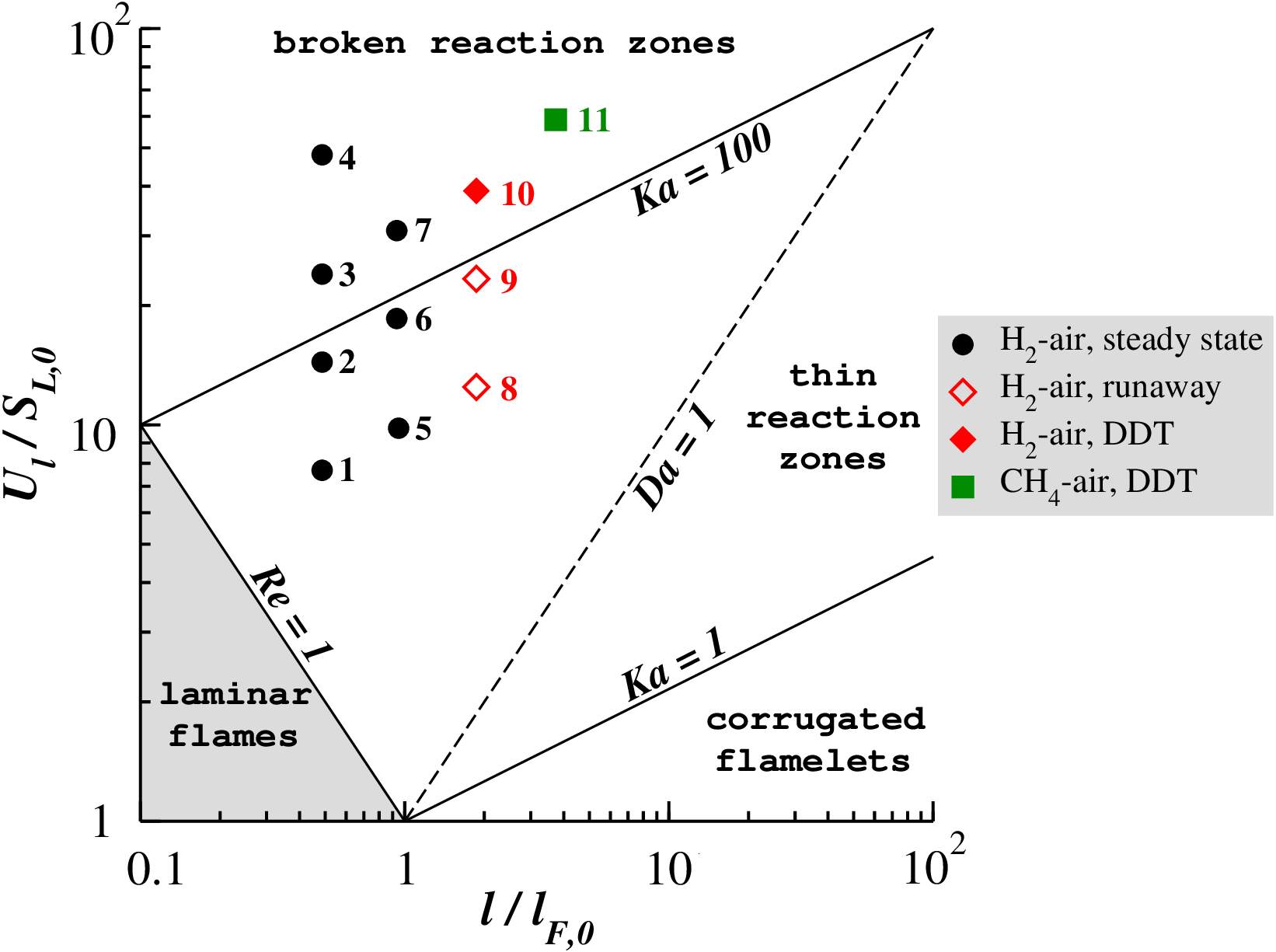}
\caption{(Color online) Combustion regime diagram \citep{Peters} showing
the simulations discussed here. Symbol color and shape indicate the
reactive mixture and the mode of burning. The full flame width
$l_{F,0} \approx 2\delta_{L,0}$ \citep{Poludnenko10}.}
\label{f:Peters}
\end{figure}

\textit{Numerical simulations.} --- Figure 1 shows a traditional combustion
regime diagram \citep{Peters} with a summary of the cases studied.
Regions of the diagram representing different burning regimes are
bounded by lines of constant nondimensional Damk\"ohler, $Da$,
Karlovitz, $Ka$, and Reynolds, $Re$, numbers \citep{Peters}. Cases 6,
7, and 10 represent several simulations testing numerical issues such
as the solution convergence and the absence of unphysical effects due
to the boundary conditions. All simulations are well-resolved with the
resolution at least $\Delta x = \delta_{L,0}/16$, where $\delta_{L,0}$
cm is the thermal width of the laminar flame in cold fuel
\citep{Poludnenko10,Poludnenko11}. Convergence was confirmed for Cases
6 \citep{Poludnenko10,Poludnenko11} and 7 using $\Delta x =
\delta_{L,0}/8 - \delta_{L,0}/32$, and convergence during the DDT
process was confirmed in Case 10 for $\Delta x = \delta_{L,0}/8 -
\delta_{L,0}/16$.

This paper focuses on Case 10 and later compares it to other
simulations shown in Fig.~1. Case 10 is a DNS of a premixed H$_2$-air
flame interacting with the high-speed, steadily driven turbulence. Its
setup is similar to the previous detailed study of Case 6
\citep{Poludnenko10,Poludnenko11}, which analyzed a steady
turbulent flame evolution in a smaller system with lower intensity
turbulence. The computational domain is a uniform
$256\times256\times4096$ Cartesian mesh with width $L = 0.518$ cm,
giving the resolution $\Delta x = \delta_{L,0}/16$ with $\delta_{L,0}
\approx 0.032$. Kinetic energy is injected at the scale $L$ to produce
homogeneous, isotropic turbulence with the characteristic velocity $U
= 1.9\times10^4$ cm/s $\approx\! 63 S_{L,0}$ at the scale $L$, where
$S_{L,0} = 3.02\times10^2$ cm/s is the laminar flame speed in cold
fuel. The large-scale eddy turnover time is $\tau_{ed} = L/U = 27.3 \
\mu$s, the integral velocity is $U_l = 1.2\times10^4$ cm/s
$\approx\!40S_{L,0}$, and the integral scale is $l = 0.12$ cm.
Resulting turbulence away from the flame has an equilibrium Kolmogorov
energy spectrum $\propto\!k^{-5/3}$ in the inertial range extending to
scales $\lesssim \delta_{L,0}$ \citep{Poludnenko10}.

\begin{figure}[!t]
\includegraphics[clip, width=0.45\textwidth]{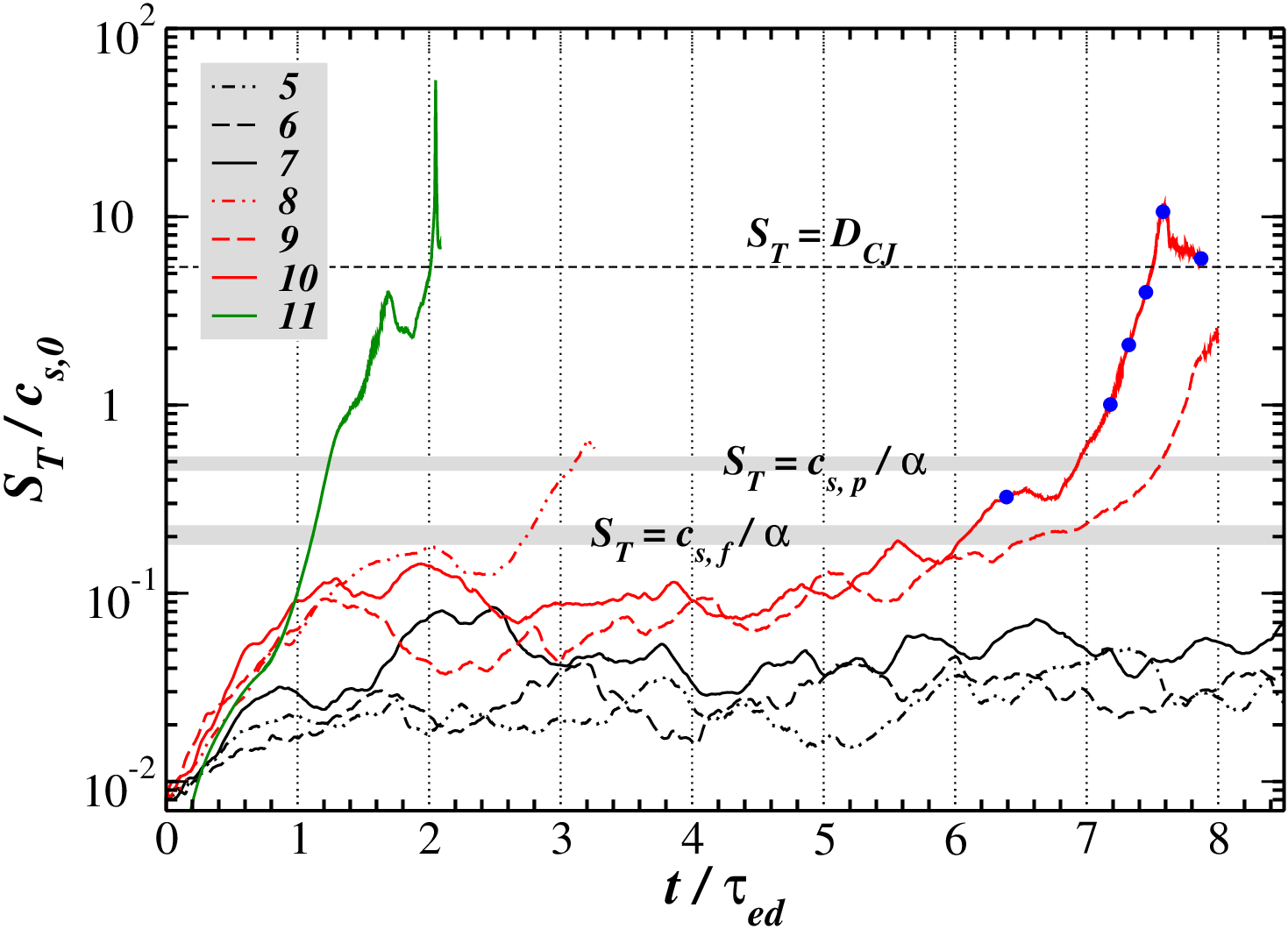}
\quad \quad
\caption{(Color online) The turbulent flame speed, $S_T$, normalized by
the sound speed in cold fuel, $c_{s,0}$. The legend gives the Case
numbers (Fig.~\ref{f:Peters}). The two shaded gray regions show the
range of critical values of $S_T$ (see \eqn{\ref{e:Criterion}}) based
on the sound speed in fuel, $c_{s,f}$, and product, $c_{s,p}$, for
fuel temperatures in the range $360-430$ K. Blue dots on the curve for
Case 10 indicate times of individual profiles in
Fig.~\ref{f:Profiles}. Time is normalized by the corresponding value
of $\tau_{ed}$ in each case.}
\label{f:Criterion}
\end{figure}

Initially, fuel has temperature $T_0 = 293$ K and pressure $P_0 =
1.01\times10^6$ erg/\cc. Steady-state turbulence is allowed to develop
for $2\tau_{ed}$. At this point ($t = 0$) a planar flame is
initialized normal to the $x$-axis. The boundary conditions are
zero-order extrapolations at the $x$-boundaries and periodic
conditions at the $y$- and $z$-boundaries. After
$\approx\!2\tau_{ed}$, the turbulent flame is fully developed and
reaches a quasi-steady state (QSS) that lasts until $t \approx
6.5\tau_{ed}$.  Figure~\ref{f:Criterion} shows the turbulent flame
speed, $S_T$, based on the fuel-consumption rate
\citep{Poludnenko10}. Turbulent flame properties during this period
are consistent with the earlier analysis of such QSS in Case 6
\citep{Poludnenko10,Poludnenko11}. In particular, the flame folded
inside the flame brush remains in the thin reaction-zone regime with
its reaction zone structure virtually unaffected by turbulence. $S_T$
is primarily controlled by the increase of the flame surface area with
an additional periodic increase $\lesssim \! 30-40\%$ due to flame
collisions and the formation of cusps.

In contrast to Case 6, the QSS in Case 10 is relatively brief
(Fig.~\ref{f:Criterion}). After $t \approx 6.5\tau_{ed}$, $S_T$ begins
to increase rapidly, becoming supersonic by $7.18\tau_{ed}$ and
exceeding the Chapman-Jouguet (CJ) detonation velocity, $D_{CJ}$, at
$7.5\tau_{ed}$. DDT occurs at $7.53\tau_{ed}$, and $S_T$ reaches its
maximum at $7.58\tau_{ed}$. At $7.63\tau_{ed}$, a fully developed
overdriven detonation emerges and quickly relaxes to $D_{CJ}$.

\begin{figure}[!t]
\includegraphics[clip, width=0.45\textwidth]{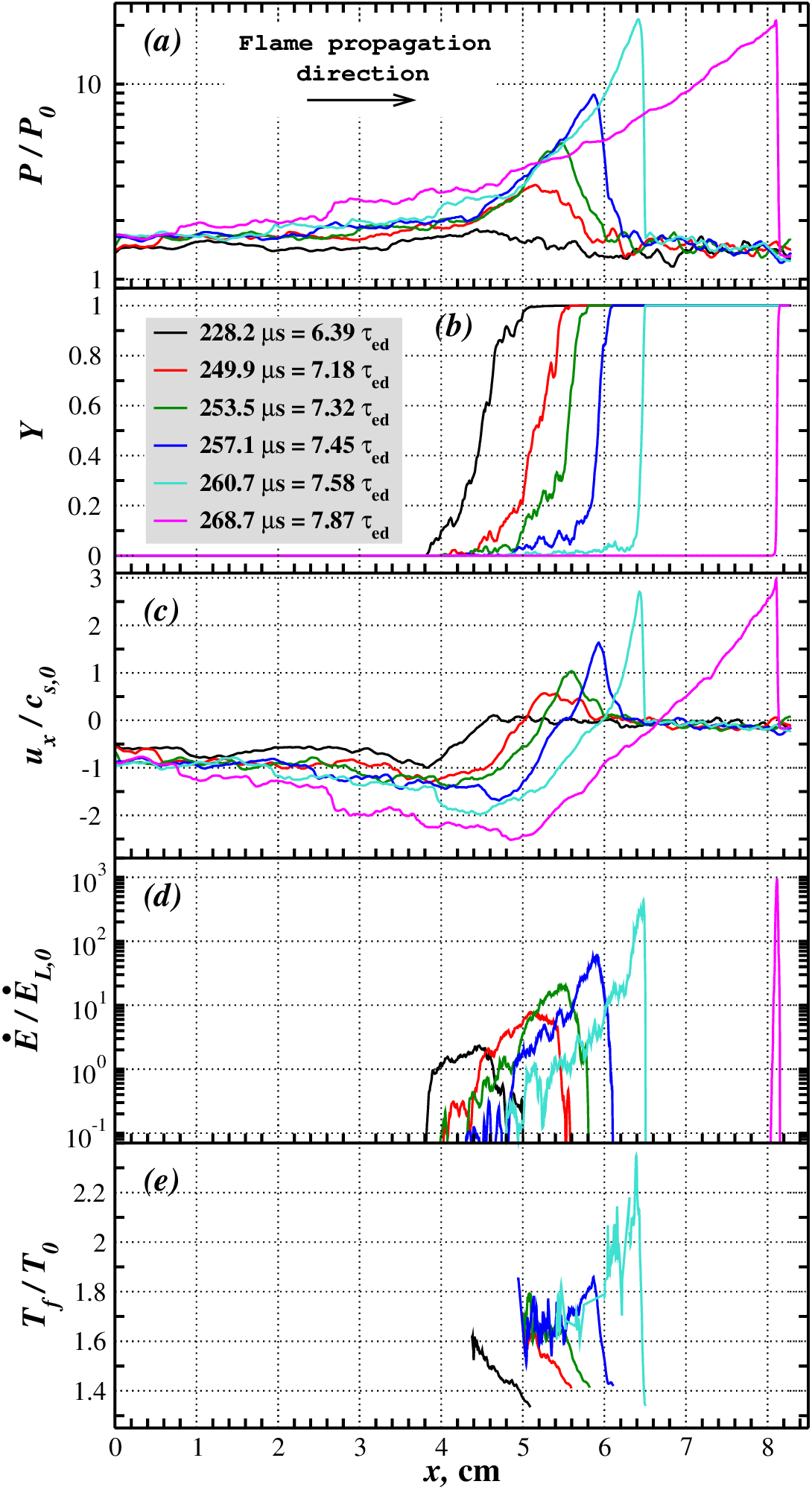}
\quad \quad
\caption{(Color online) The $y$-$z$-averaged profiles of (a) pressure,
$P$, (b) fuel mass fraction, $Y$, (c) $x$-velocity, $u_x$, (d)
energy-generation rate per unit volume, $\dot{E}$, and (e) temperature
of pure fuel ($Y \ge 0.95$) in Case 10. The time from ignition for
each profile is shown in panel (b) and indicated with blue dots in
Fig.~\ref{f:Criterion}. $\dot{E}$ is normalized by its value in a
planar laminar flame propagating in cold fuel, $\dot{E}_{L,0} =
qS_{L,0}\rho_0/\delta_{L,0}$, where $q$ is the chemical energy release
and $\rho_0$ is the density of cold fuel. $T_f$ is shown inside the
flame brush up to the moment of detonation formation ($t \le
7.58\tau_{ed}$).}
\label{f:Profiles}
\end{figure}

The system evolution during this process is shown in
Fig.~\ref{f:Profiles}. At $6.39\tau_{ed}$, a slight overpressure
arises inside the flame brush, but the energy-generation rate per unit
volume, $\dot{E}$, is still close to its value in the planar laminar
flame. As the pressure grows and the turbulent flame accelerates, fuel
inside the flame brush is compressed and heated. This increases the
local flame speed, $S_L$, causing $\dot{E}$ to rise. At later times,
$\dot{E}$ exceeds the laminar value by $\sim\!2$ orders of magnitude.
Such accelerated burning leads to further fuel compression and larger
$S_L$. The resulting feedback loop drives a catastrophic runaway
process that produces a large pressure build-up and creates strong
shocks inside the flame brush. These, in turn, create conditions in
which a detonation can arise. (Details of this last stage will be
presented in a separate paper.)

Up until the moment of DDT, the average fuel temperature, $T_f$,
inside the flame brush remains $<\!700$ K (Fig.~\ref{f:Profiles}e),
and the corresponding induction times are much larger than all
dynamical timescales. At all times, the average internal flame
structure (reconstructed using method described in
\citep{Poludnenko10}) is close to that of a laminar flame in fuel with
the corresponding $T_f$ and pressure. Thus, during the runaway,
burning is controlled by flame propagation and not by autoignition,
which precludes the formation of global spontaneous reaction waves.

\textit{Mechanism of the spontaneous runaway. }--- Consider an
unconfined fluid volume $V$ with the total internal energy
$\varepsilon$. To increase the pressure inside $V$ (as in
Fig.~\ref{f:Profiles}), an energetic process must generate energy
comparable to $\varepsilon$ on the characteristic sound-crossing time
of this volume, i.e., $\dot{\varepsilon} \sim \varepsilon/t_s$. If
this volume represents a flame with width $\delta$ and cross-sectional
area $L^2$, i.e., $V = \delta L^2$, then the burning speed of the
flame is defined as $S = \dot{m}/\rho_fL^2$, where $\dot{m} =
\dot{\varepsilon}/q$ is the total fuel-consumption rate and $\rho_f$
is the fuel density. Then the condition $\dot{\varepsilon} \sim
\varepsilon/t_s$ can be rewritten as $S \sim c_sE/q\rho_f$, where $t_s
= \delta/c_s$, $c_s$ is the sound speed, and $E = \varepsilon/V$ is
the internal energy per unit volume. The flame here may be laminar,
turbulent, or distributed, provided it has the required burning speed.

In order to examine the physical meaning of this condition on $S$,
assume an ideal gas equation of state, $E = P/(\gamma - 1)$. At the
start of the runaway, pressure is nearly constant across the
flame. Then the product density is $\rho_p = \rho_fT_f/T_p =
\rho_fT_f/(T_f + q/C_p) = P/(P/\rho_f + q(\gamma-1)/\gamma)$, where
$T_p$ is the product temperature and $C_p$ is the specific heat at
constant pressure. For energetic reactive mixtures, the denominator
$P/\rho_f + q(\gamma-1)/\gamma$ can be approximated as
$q(\gamma-1)$. Here $q = 43.28RT_0/M \gg P_0/\rho_0$
\citep{Poludnenko10} and at the onset of the runaway, $P \approx
1.5P_0$ and $\rho_f \approx \rho_0$, giving the accuracy of this
approximation $\approx\!6\%$. Thus, $\rho_p
\approx P/q(\gamma-1)$, and
\beq
S \sim \frac{c_s}{q\rho_f}E = \frac{c_s}{\rho_f}\frac{P}{q(\gamma-1)}
\approx \frac{c_s}{\alpha} \equiv S_{CJ},
\label{e:Criterion}
\eeq
where $\alpha = \rho_f/\rho_p$ is the fluid expansion factor.

In the reference frame of a steady flame, $\rho_pU_p = \rho_fU_f =
\rho_fS$, where $U_f$ and $U_p$ are the velocities of the
fuel and product, respectively. Thus, \eqn{\ref{e:Criterion}} is
equivalent to the statement that $U_p = c_s$. If $c_s$ is taken as the
sound speed in the product, then the flame with the speed satisfying
eq.~(\ref{e:Criterion}) is a CJ deflagration \citep{Williams}.

The speed of a CJ deflagration, $S_{CJ}$, is a theoretical maximum
speed of a steady-state flame. The discussion above shows that such a
flame generates enough energy on a sound-crossing time to raise its
internal pressure and, thus, disrupt its steady-state structure. Real
laminar flames, both chemical \citep{Williams} and thermonuclear
\citep{Khokhlov97ApJ}, do not have burning speeds that approach
$S_{CJ}$. Turbulent flames, however, can develop such high values of
$S_T$.

Unlike a laminar flame, in which the local sound speed increases
smoothly from its value in the fuel, $c_{s,f}$, to that in the
product, $c_{s,p}$, a turbulent flame effectively consists of two
fluids with either $c_{s,f}$ or $c_{s,p}$. Figure~\ref{f:Criterion}
shows $S_{CJ}$ based on both $c_{s,f}$ and $c_{s,p}$. Dissipative
heating of fuel by turbulence causes $c_{s,f}$ and $c_{s,p}$ to
increase and $\alpha$ to decrease. Thus, the horizontal shaded gray
areas show the range of values of $S_{CJ}$ corresponding to fuel
temperatures $\approx\!360-430$ K. In particular, in Case 10, $T_f
\approx 360$ K at $2\tau_{ed}$ (lower bound of the shaded regions) and
increases to $\approx \! 430$ K by $6.5\tau_{ed}$ (upper bound).

Figure~\ref{f:Criterion} shows that, upon first reaching the QSS,
$S_T$ is close to, but still below, $c_{s,f}/\alpha$, which prevents
the onset of the runaway. During the time $(2 - 6.5)\tau_{ed}$,
turbulent heating of fuel increases $S_L$ by a factor of $\approx \!
2$, thus accelerating $S_T$ above the critical value $c_{s,f}/\alpha$
and allowing the runaway to begin.  Figure~\ref{f:Profiles}(c) shows
that, at this point, the product velocity indeed becomes
$\approx\!c_{s,f}$. Furthermore, the growth rate of $S_T$ increases
significantly once $S_T$ becomes $> \!  c_{s,p}/\alpha$, i.e., when
$U_p$ becomes supersonic relative to both sound speeds. Note also that
the transition from a QSS to a detonation occurs on a sound-crossing
time of the turbulent flame $t_{s} = \delta_T/c_{s,0} \approx 27 \mu$s
$\approx \tau_{ed}$, where $\delta_T \approx 1$ cm is the flame-brush
width (Fig.~\ref{f:Profiles}b) and $c_{s,0} \approx 3.7\times10^4$
cm/s.

Figure~\ref{f:Criterion} also shows $S_T$ for turbulent H$_2$-air
flames for other values of $U_l$ and $l$. In Cases 5-7, $S_T$ remains
well below $c_{s,f}/\alpha$, and the flame evolves in the QSS, as
described in \citep{Poludnenko10,Poludnenko11}. This QSS was observed
over significantly longer periods of time than shown in
Fig.~\ref{f:Criterion}, e.g., $16\tau_{ed}$ in Case 6. Cases 1-4 were
similar and so are not shown. The runaway process was also observed in
Cases 8 and 9, in which, however, the flame accelerated quickly and
left the domain before DDT could occur. Note that the overall growth
rate of $S_T$ in Cases 8 and 9 was lower than in Case 10 ($\tau_{ed}$
increases with decreasing $U_l$).

To determine the dependence of the results on the reaction model, we
carried out a similar simulation for a stoichiometric CH$_4$-air
mixture. In this case, $\delta_{L,0} = 0.042$ cm is close to that in
H$_2$-air, but $S_{L,0} = 38$ cm/s is eight times lower
\citep{Kessler10}. The CH$_4$-air system also showed DDT, but at a higher
turbulent intensity relative to $S_L$ ($U_l = 2.24\times10^3$ cm/s
$\approx 59S_L$) and in a larger system ($l = 0.31$ cm, $L = 1.328$
cm) (Case 11, Figs.~\ref{f:Peters} and \ref{f:Criterion}). The overall
evolution, however, was different from Case 10. The time to DDT was
$\approx \! 2\tau_{ed}$, and the flame never developed a QSS. The
flame accelerated significantly relative to fuel, which required a
longer domain to observe DDT, and, in contrast with Case 10, a strong
well-defined global shock formed and ran ahead of the flame brush.

The key aspect of the spontaneous DDT mechanism discussed here is that
it does not place any specific constraints on the equation of state,
reaction model, or the flame properties. A decrease of fluid density
with increasing temperature in an exothermic process means that, at a
high but subsonic burning speed, the flow of products becomes
supersonic relative to the flame, irrespective of how burning
occurs. This ensures that the pressure wave remains coupled to the
region in which the energy release occurs (note the location of peaks
of $P$ and $\dot{E}$ in Fig.~\ref{f:Profiles}b and c). This is in
contrast with the spontaneous reaction-wave model \citep{Zeldovich},
which requires very specific hot-spot properties in order for the
reaction wave and the pressure pulse it produces to remain properly
coupled.

Figure~\ref{f:Peters} suggests that there is both a minimal system
size and a minimal relative turbulent intensity at which DDT is
possible, and they appear to increase for reactive mixtures with
slower laminar flames. Applying \eqn{\ref{e:Criterion}} to establish
whether DDT can occur depends on our ability to predict the turbulent
flame speed for given $U_l$ and $l$. This is particularly difficult in
the high-speed regimes where spontaneous DDT is most likely to occur.
Further studies using more detailed chemical kinetics models are
required to establish the range of regimes in which DDT is to be
expected for realistic reactive mixtures and to investigate the
possibility of flame extinction in the presence of high-intensity
turbulence.

\begin{acknowledgments}
We thank Vadim Gamezo, Craig Wheeler, and Forman Williams for
valuable discussions. This work was supported by the AFOSR grant
F1ATA09114G005 and by the ONR/NRL 6.1 Base Program.
\end{acknowledgments}


\providecommand{\noopsort}[1]{}\providecommand{\singleletter}[1]{#1}%

\end{document}